# High-performance parallel classical scheme for simulating shallow quantum circuits

Shihao Zhang[1*,3], Jiacheng Bao[2*], Yifan Sun[1], Lvzhou Li[3], Houjun Sun[2$], and Xiangdong Zhang[1+]

[1] Key Laboratory of advanced optoelectronic quantum architecture and measurements of Ministry of Education, Beijing Key Laboratory of Nanophotonics & Ultrafine Optoelectronic Systems, School of Physics, Beijing Institute of Technology, 100081, Beijing, China

[2] Beijing Key Laboratory of Millimeter wave and Terahertz Techniques, School of Information and Electronics, Beijing Institute of Technology, Beijing 100081, China

[3] School of Computer Science and Engineering, Sun Yat-Sen University, Guangzhou 510006, China

* These authors contributed equally to this work.

[+$] Author to whom any correspondence should be addressed.

**E-mail**: zhangxd@bit.edu.cn and sunhoujun@bit.edu.cn

## Abstract

Recently, constant-depth quantum circuits are proved more powerful than their classical counterparts at solving certain problems, e.g., the two-dimensional (2D) hidden linear function (HLF) problem regarding a symmetric binary matrix. To further investigate the boundary between classical and quantum computing models, in this work we propose a high-performance two-stage classical scheme to solve a full-sampling variant of the 2D HLF problem, which combines traditional classical parallel algorithms and a gate-based classical circuit model together for exactly simulating the target shallow quantum circuits. Under reasonable parameter assumptions, a theoretical analysis reveals our classical simulator consumes less runtime than that of near-term quantum processors for most problem instances. Furthermore, we demonstrate the typical all-connected 2D grid instances by moderate FPGA circuits, and show our designed parallel scheme is a practically scalable, high-efficient and operationally convenient tool for simulating and verifying graph-state circuits performed by current quantum hardware.

## 1. Introduction

Quantum computers have been the focus of numerous studies and are expected to play an important role in future information processing, since they can outperform classical computers in certain computational tasks [1-5]. Recently, the concept of "quantum supremacy/advantage" has been proposed, involving some models [6,7] and being demonstrated experimentally in different tasks [8,9]. At the same time, an unconditional "quantum advantage" has also been theoretically proved in the quantum shallow circuits for solving the 2D hidden linear function (HLF) problem [10], which provides a separation between the capabilities of quantum and conventional classical circuits.

The HLF problem is defined over a given $n \times n$ symmetric binary matrix $A$ and the task is to find a binary vector $z \in \{0,1\}^n$ satisfying $q(x) = 2z^T x \pmod 4$ for all $x \in \text{Ker}(A)$, where $q(x) = x^T A x \pmod 4$ and $\text{Ker}(A) = \{x \in \{0,1\}^n : Ax = 0 \bmod(2)\}$ is the binary null-space of $A$. Specifically, an instance of the 2D HLF problem can be alternatively described by a square grid graph $G(A)$ with $n = N \times N$ vertices as depicted in figure 1(a), where $A$ acts as the adjacency matrix of $G(A)$. By definition the connections of edges in the graph $G(A)$ are determined by off-diagonal part of $A$, and the $i$th vertex of $G(A)$ is marked yellow (blue) if the value of the diagonal element $A_{i,i}$ is 1 (0). Thus, the basic idea of an advanced quantum solution [10] is to prepare a $n$-qubit entangled graph state corresponding to $G(A)$ by constant-depth quantum circuits $Q_N$ and then measure each qubit in Pauli $X/Y$ bases indicated by the color of each vertex, which can output a random $n$-bit string as a solution to the 2D HLF problem (see appendix A). As comparison, Bravyi *et al*. first show that no sub-logarithmic-depth classical circuit $C_N$ with bounded fan-in can solve all instances of sufficiently large 2D HLF problems with high probability, and such a quantum advantage stems from the nonlocal quantum correlations [10].

Although a separation is established between the depth of quantum and classical circuits for outputting one solution to the 2D HLF problem, its intriguing variant aimed to obtain all different solutions to any problem instance is short of study. This can be viewed as a full-sampling version of the original problem and thus named the FS2D HLF problem for clarity. In recent years, some sampling problems have raised great interest due to their appealing roles in the computational complexity theory and quantum computing applications [8,9,11-18], and high-performance classical simulations for tasks of random quantum circuits sampling [19-21] and Boson sampling [22-24] have been developed to promote the design and optimization of near-term quantum sampling experiments. In this direction, the study of effective classical simulators for fully sampling a constant-depth quantum circuit family $\{Q_N\}$ related to FS2D HLF problems may bring new discoveries. The target quantum protocol here is to run $Q_N$ enough times and perform the particular projective measurements on the output state in each run, like those practically performed experiments to obtain outcome distributions [8,25-29]. In fact, various existing classical methods that are capable of simulating the so-called stabilizer (Clifford) circuits have extended from the growing stabilizer formalism [30-33] to recent tensor-network contraction algorithms [34], yet most of these works pay less attention to the parallel diagram for the classical circuit design.

Motivated by these investigations, in this paper we propose a time-efficient and scalable classical scheme for any FS2D HLF problem instance in two stages as shown in figures. 1(b) and 1(c). The first stage is a CLA module to compute some linear-algebraic quantities about the given matrix $A$ (e.g. its binary rank $r$) within time $O(\log^2 n)$, followed by utilizing these quantities as guides to repeatedly run a local classical parallel circuit (CPC) that can output all $k$ different solutions one by one in the second stage. Note the CPC is designed according to an analogy to the shallow quantum circuits and thus owns a constant depth. In the following section we can prove $k = 2^r$. Thus if $r = n$, then $k = 2^n$ and all $n$-bit strings $z \in \{0,1\}^n$ are solutions. If $r < n$, then $k \leq 2^{n-1}$ and the expected number of repetitive run-and-measure quantum procedures is theoretically $\Theta(k \log k)$ for uniformly superposed stabilizer states as the known "coupon collector problem" [35], i.e., such a quantum protocol about $Q_N$ requires time $T_Q = \Theta(r 2^r)$. As an exact classical simulator for solving FS2D HLF problems, our two-stage scheme has total runtime scaling as $T_C = O(\log^2 n) + \Theta(2^r)$ as comparable to $T_Q$.

This paper is organized as follows. We begin by describing the design ideas and working principles of our two-stage simulation scheme in section 2. In section 3 we give the theoretical analysis of this scheme's performance in terms of its total time and memory cost, which can exhibit high efficiency and scalability for simulating target shallow circuits implemented on near-term quantum hardware, and then present the experimental testing for several typical instances on a practical FPGA platform in section 4. Finally, we discuss the underlying causes behind such high-performance of our scheme and conclude the paper in section 5.

## 2. The two-stage classical simulation scheme

At first, we describe the whole framework of our two-stage simulation scheme. Here we write

$L_q = \text{Ker}(A)$ and let $L_q^\perp$ be the orthogonal complement of $L_q$. For the FS2D HLF problem specified by $A$, it has been pointed out that once a solution denoted by $z^a = z_1^a z_2^a \ldots z_n^a$ is obtained, then $z^a \oplus y$ is also a solution if and only if $y \in L_q^\perp$, and thus the number of different solutions is $k = |L_q^\perp|$ (see supplementary material of Ref. [10] for proof). Here the symbol $\oplus$ denotes addition of binary strings modulo two, i.e., the bitwise XOR (exclusive-OR) operation. In this way, the question turns into how to find an arbitrary solution $z^a$ as well as all element vectors $y$'s in $L_q^\perp$. We demonstrate a two-stage classical scheme including function modules depicted in figures. 1(b) and 1(c) for divided tasks, which combine together to finally output all correct solutions.

In figure 1(b), a series of linear algebraic quantities about the given binary matrix $A$ have been parallel computed in the CLA module, including: (1) computing a maximal linearly independent set of column vectors $L_{\text{ind}} = \{l_{p_1}, l_{p_2}, \ldots, l_{p_r}\}$ of $A$, where the subscripts $P = \{p_1, p_2, \ldots, p_r\}$ are the set of pivot positions and $r$ is the binary rank; (2) finding a basis set for the null-space $L_q$; (3) doing the matrix-vector multiplication $q(x) = x^T A x \pmod 4$ for all $x \in L_q$; (4) computing one solution $z^a = z_1^a z_2^a \ldots z_n^a$ of linear equations $2z^T x \pmod 4 = q(x)$ over $z \in \{0,1\}^n$. All these procedures over the binary field can be realized by the $O(\log^2 n)$-time parallel algorithms using a polynomial number of classical processors [36,37].

Next, we proceed to the second stage to produce all element vectors $y$'s in $L_q^\perp$. Using the identity $\text{Ker}(A)^\perp = \text{Row}(A) = \text{Col}(A)$ for the row and column spaces of binary symmetric matrix $A$ [38,39], we find that $\text{Col}(A)$ spanned by the set $L_{\text{ind}}$ equals to $L_q^\perp$. Therefore, the total number of solutions is $k = |L_q^\perp| = |\text{Col}(A)| = 2^r$, and thus the question is how to explicitly express all element vectors in $\text{Col}(A)$ by certain depth-efficient classical means. We address this issue by proposed constant-depth classical parallel circuit in figure 1(c), and illustrate its inner structure and working principles.

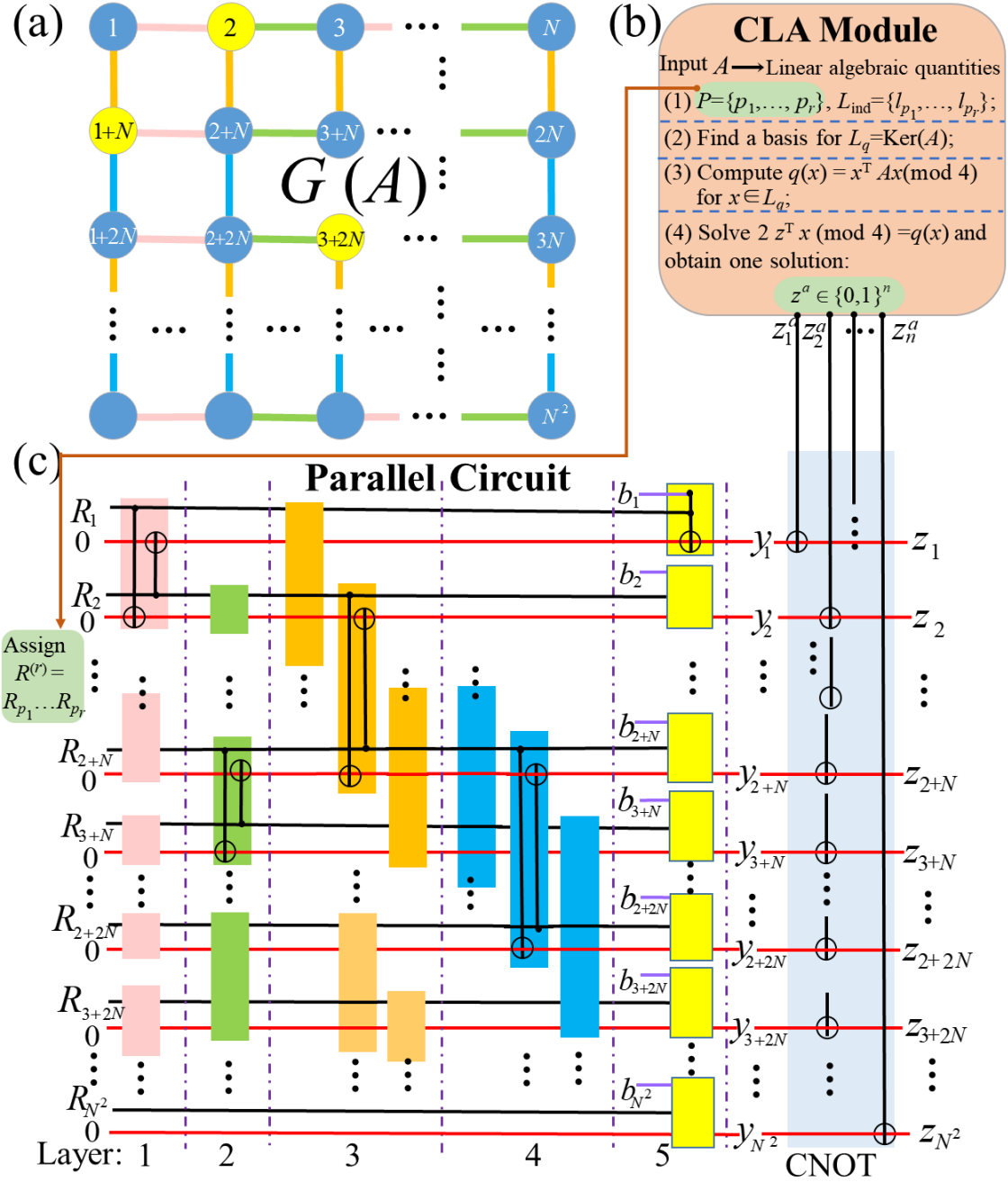

**Figure 1.** Classical scheme to solve the FS2D HLF problem. (a) Graph $G(A)$ of the square grid with $n = N \times N$ vertices determined by the given symmetric binary matrix $A$, where blue and yellow vertices correspond to the diagonal element $A_{i,i} = 0$ and $A_{i,i} = 1$, respectively. Edges connecting each pair of vertices are marked with different colors (pink, green, orange, blue). Here the depicted graph is a class of typical all-connected 2D grids. (b) The CLA module that computes required linear algebraic quantities of $A$ and one solution $z^a \in \{0,1\}^n$ to the original 2D HLF problem. (c) Classical parallel circuit with a constant depth. The $n$ groups of input bits carried by black wires and red wires are denoted by $(R_i, y_i)$ ($i = 1, 2, \ldots, n$) with initial values ($R_i, y_i = 0$), and the pivot position set $P = \{p_1, p_2, \ldots, p_r\}$ from (b) is used

to assign the input substring to traverse over $\{0,1\}^r$ with rest $R_i$'s fixed to 0, thus the corresponding $2^r$ final output bit strings $z$'s after the final CNOT-gates layer controlled by $z^a = z_1^a z_2^a \ldots z_n^a$ would be all solutions to the problem.

The parallel circuit shown in figure 1(c) is designed according to the structure of the 2D grid graph $G(A)$ in figure 1(a). The initial input of parallel circuit includes $n$ pairs of bits $(R_i, y_i = 0)$ ($i = 1, 2, \ldots, n$), each of which are carried by the black and red wires respectively. Next, the arrangement of the operation units indicated by pink, green, orange and blue rectangles for processing certain two groups of bits $(R_i, y_i)$ and $(R_j, y_j)$ in the first four layers of the parallel circuit is determined by the connection of edges $(i, j)$ with the same color in the 2D grid graph. For example, the pink rectangle operation unit (ROU) acting on $(R_1, 0)$ and $(R_2, 0)$ in the first layer corresponds to the pink edge connecting vertices 1 and 2 in figure 1(a), and the cases for ROUs in the second, third and fourth layers are displayed in a similar way. For the initial input $(R_i, y_i = 0)$ ($i = 1, 2, \ldots, n$), after a ROU containing a pair of classical CNOT gates acting on $(R_i, y_i)$ and $(R_j, y_j)$ respectively, the outputs can be obtained by $y_i \to y_i \oplus R_j$ and $y_j \to y_j \oplus R_i$ as demonstrated in Ref. [40]. Therefore, after 4 layers of ROUs the $i$th output bit in the red wire can be given by

$$y_i = \bigoplus_{j \in N(i)} R_j \quad (i = 1, 2, \ldots, n), \tag{1}$$

where $N(i)$ is the set of nearest-neighbor vertices of vertex $i$.

For the next gate operations in the 5th layer, an additional input control bit $b_i \equiv A_{i,i}$ (marked by a purple wire) for the diagonal elements of $A$ needs to be added to each group of bits $(R_i, y_i)$. Applying the classical Toffoli gate represented by yellow square unit to the set of bits $(b_i, R_i, y_i)$ as shown in figure 1(c) would update the third bit as $y_i \to y_i \oplus b_i R_i$ [4]. At this point, the outcome carried by the $i$th red wire after 5 layers of operation units changes into

$$y_i = (\bigoplus_{j \in N(i)} R_j) \oplus b_i R_i \quad (i = 1, 2, \ldots, n) \tag{2}$$

as the $i$th bit of the output string $y = y_1 y_2 \ldots y_n$.

Note that for any $i$, one has $A_{i,j} = 1$ for $j \in N(i)$, $A_{i,j} = 0$ for $j \notin N(i)$, and the input diagonal bit denoted $A_{i,i} \equiv b_i$. Thus, each output bit $y_i$ in equation (2) can be equivalently expressed by the

elements of $A$ as

$$y_i = \bigoplus_{j=1}^{n} A_{i,j} R_j \ (i=1,2,\ldots,n), \tag{3}$$

which implies that the binary matrix-vector multiplication $y = AR$ can be realized exactly by inputting the string $R = R_1 R_2 \ldots R_n$ into this classical parallel circuit.

As mentioned above, we denote the maximal linearly independent set of column vectors of $A$ as $L_{\text{ind}} = \{l_{p_1}, l_{p_2}, \ldots, l_{p_r}\}$, where the subscripts $S_p = \{p_1, p_2, \ldots, p_r\}$ are the set of pivot positions and $r$ is the matrix rank over the binary field. Thus, by definition the binary matrix column space $\text{Col}(A)$ is spanned by the vectors in $L_{\text{ind}}$ and has dimension $r$. As for the corresponding classical parallel circuit, when we mark a $r$-bit substring of input $R = R_1 R_2 \ldots R_n$ with subscripts in the set $P = \{p_1, p_2, \ldots, p_r\}$ as $R^{(r)} = R_{p_1} R_{p_2} \ldots R_{p_r}$ and fix the rest $(n-r)$ input $R_i$'s with subscripts in the set $[1,n] \setminus P$ to 0, then using the observation that the column vector in $L_{\text{ind}}$ is $l_{p_j} = [A_{1,j}, A_{2,j}, \ldots, A_{n,j}]^T$ together with equation (3) we can infer: the vector $y = y_1 y_2 \ldots y_n$ in equation (2) before the last CNOT-gates layer would equal to $\bigoplus_{i=1}^{r} R_{p_i} l_{p_i}$ ( $p_i \in P$, $l_{p_i} \in L_{\text{ind}}$ ). Finally, a layer of CNOT gates with the $i$th bit of $z^a = z_1^a z_2^a \ldots z_n^a$ as the control bit from CLA module are applied to the vector $y = y_1 y_2 \ldots y_n$.

In this setting, we let the pivot position set $P = \{p_1, p_2, \ldots, p_r\}$ computed from CLA module as input information (indicated by brown polyline) to control the running of parallel circuit in figure 1(c), which can produce all $2^r$ solutions to the FS2D HLF problem. Specifically speaking, when we traverse the values of those input substrings $R^{(r)} = R_{p_1} R_{p_2} \ldots R_{p_r}$ over $\{0,1\}^r$ and fix the values of the rest $(n-r)$ input $R_i$'s to 0, the $2^r$ vectors $y$'s denoted by red wires before the final CNOT-gate layer are exactly the linear combinations of basis vectors spanning the column space $\text{Col}(A)$. Then, $2^r$ different solutions denoted $z$'s to the FS2D HLF problem can be obtained one by one through $n$ parallel CNOT gates as $z = C^{z^a} \text{NOT}(y) = y \oplus z^a \in \{0,1\}^n$ during the traversal of $R^{(r)} \in \{0,1\}^r$.

## 3. Theoretical analysis for the performance

In this section, we theoretically analyze the time and memory cost for simulating the quantum proposal by our classical scheme. Since the basic components used in our classical parallel circuit are in one-to-one correspondence with those in the quantum circuit $Q_N$ in figure 4 of appendix A, it also has a constant depth and the number of basic elements increases as $O(n)$ with the problem size $n$. Therefore, our two-stage classical scheme runs in total time $T_C = O(\log^2 n) + \Theta(2^r)$ using a polynomially bounded number of processors, while repeating the quantum circuit $Q_N$ consumes time as $T_Q = \Theta(r 2^r)$. Besides, for the special full-rank instances with $r = n$, all $2^n$ bit-strings over $\{0,1\}^n$ are definitely outcomes. To evaluate the performance of our simulation scheme in a more realistic way, here we estimate its actual runtime under reasonable assumptions considering the capabilities of current classical and quantum computing hardware.

At first, the actual runtimes of the CLA module and the constant-depth classical parallel circuit are denoted as $T_1^{act} = c_1 \times \log^2 n$ and $T_2^{act} = c_2 \times 2^r$, respectively. Together, the total runtime of this two-stage classical scheme to generate all solutions is $T_C^{act} = T_1^{act} + T_2^{act} = c_1 \times \log^2 n + c_2 \times 2^r$, where the coefficients $c_1$ and $c_2$ are determined by the performance of used classical computing devices, e.g., the speed of basic logical gates. In contrast, the actual runtime for reliably repeating the corresponding quantum circuit is $T_Q^{act} = c_3 \times r 2^r$, where $c_3$ is associated with technical conditions on quantum operation speeds. Then, the ratio between $T_C^{act}$ and $T_Q^{act}$ is a natural metric for characterizing the performance (efficiency) of our classical simulator aimed at a quantum hardware as:

$$\frac{T_C^{act}}{T_Q^{act}} = \frac{c_1}{c_3} \frac{\log^2 n}{r 2^r} + \frac{c_2}{c_3} \frac{1}{r}. \tag{4}$$

The result $T_C^{act} / T_Q^{act} \leq 1$ implies the designed classical simulator consumes less time than the target quantum experiment, and a smaller ratio value indicates a more efficient classical simulation. Equation (4) shows $T_C^{act} / T_Q^{act}$ decreases as $r$ increases for the given $n$ and coefficients $c_1$, $c_2$, $c_3$, indicating $T_C^{act} \leq T_Q^{act}$ may appear when the value of $r$ is greater than a certain lower bound.

Here we make some reasonable parameter assumptions for illustration. At first, we roughly assume the coefficients $c_1$, $c_2$ and $c_3$ are similar in view of moderate classical gate speeds and progressive quantum technology [41,42], e.g., gate time of tens of nanoseconds. Then the relation $\log^2 n \leq 2^r$ would lead to $T_C^{\text{act}}/T_Q^{\text{act}} \leq 1$ assuming $c_1 = c_2 = c_3$ and $r \geq 2$ for any given $n$ in equation (4), which can be expressed as

$$r \geq r_0 = \lceil 2\log(\log n) \rceil. \tag{5}$$

Considering the gate speed of current quantum devices actually cannot catch up with sophisticated classical (super)computers [19-23], we suppose $c_1/c_3 < 1$ and $c_2/c_3 < 1$ in equation (4) and thus the lower bound $r_0$ in equation (5) ensures $T_C^{\text{act}}/T_Q^{\text{act}} \leq 1$ for practical cases. This LogLog $r \sim n$ relation reveals that even for a quite large $n$, e.g., $n = 10^6$ corresponding to millions of qubits, the binary rank satisfying $r \geq r_0 = 9$ is sufficient to ensure $T_C^{\text{act}} \leq T_Q^{\text{act}}$ and then the ratio $T_C^{\text{act}}/T_Q^{\text{act}}$ in equation (4) behaves as $1/r$ asymptotically (up to constant factors) when $r$ increases from $r_0$ to $n$. At this point, the excellent performance of our classical scheme for instances with $r \in [r_0, n]$ shows it can serve as a scalable and efficient classical sampler for a wide range of such shallow quantum circuits.

To highlight such traits, here we particularly focus on the study of FS2D HLF problems defined on a typical class of all-connected 2D grid graphs as shown in figure 1(a), i.e. each vertex reaches its full degree in the square grids. By proving that the binary rank $r$ of such a size-$n$ adjacency matrix must lie in the interval $r \in [n - \sqrt{n}, n]$ (see appendix B), our classical scheme indicates an extraordinary time-efficient simulator as the ratio $T_C^{\text{act}}/T_Q^{\text{act}} \ll 1$ for sufficiently large $n$. In the following, we test the above theory on moderate all-connected 2D instances by demonstrative experiments on the FPGA platform.

## 4. Experimental testing for FS2D HLF problem

Above design can be effectively demonstrated on an FPGA platform due to abundant logical resources provided for many kinds of classical circuits. First, the pivot position set $P$ and any one solution $z^a$ from the CLA module in figure 1(b) can be calculated before the FPGA experiment. Then, we implement the stage of parallel circuits with different sizes on the FPGA platform by a code design. Currently, the scale of genuine multipartite entangled states generated in quantum experiments is up to about 20 qubits [28,29,43]. To compare with the candidate quantum experiments, we consider an FPGA chip with 25 channels. Thus, the experimental results for the cases with $n = 2\times 2$, $3\times 3$, $4\times 4$ and $5\times 5$ are provided here.

In the experiments, we choose the FPGA product of Xilinx company for development. The specific device is XC7K410TFFG900, and ISE14.4 software development environment is adopted. As for the implementation method, we use Verilog HDL hardware description language to do the code design at the register transfer level (RTL). In order to increase the clock rate for a high processing speed and a sufficient resolution of timing closure, we use the pipeline technology to optimize the design module. Because the parallel circuits can be divided into several processing units (2×2: 3 layers; 3×3/4×4/5×5: 5 layers), we use each layer of processing unit as a primary pipeline processing unit, and insert registers between layers to store temporary data. According to this arrangement, the whole module is divided into 3/5 levels of pipeline processing so that the clock rate can be effectively increased.

As an example, in figure 2(a) we show the schematic diagram of FPGA at RTL for solving an instance of $n = 2\times 2$ ($b = 0000$), and the corresponding parallel circuit model is given by figure 2(b). Some devices used in the intermediate process are listed in the detailed insets (i)-(vii) of figure 2. By comparison, it can be seen that the compiled circuit basically retains the topological relationship between layers in figure 2(b). For the external interface design, we arrange all the input signals in figure 2(a) on the leftmost side, where I_rst is the module reset signal and I_sysclk is running clock of the module. The module X_12 (CONSTANT REG) and 4 LATCHes ( $y^0[1]$, $y^0[2]$, $y^0[3]$ and $y^0[4]$ ) work together to provide the 4-bit input 0000 for the parallel circuit, which corresponds to the 4-bit '0' input at the leftmost side in figure 2 (b). The 2-bit adder module ADD_2u with one input '01' and the FLIP-FLOP module $\mathbf{R}_C^0$ work together to traverse $R_1R_2 \in \{0,1\}^2$ according to $P = \{1,2\}$ here, and then the output bits of $\mathbf{R}_C^0$ as $R_1R_2$ are combined with 2-bit 'ground line', which represents '00' for $R_3R_4$ so as to form a complete 4-bit input signal for $\mathbf{R}^0$. This signal modulation part corresponds to the input $R_1R_2R_3R_4$ and the guidance information $P$ at the leftmost side in figure 2 (b). The input signals b[1:4] and Za[1:4] correspond to the input string $b = b_1b_2b_3b_4$ and $z^a = z_1^a z_2^a z_3^a z_4^a$ in figure 2(b), respectively. In the intermediate process, the pink, orange and blue ADDER blocks correspond to those CNOT gates enveloped by the box with the same color in figure 2(b). The combination of 4 AND GATEs and 4 ADDERs in yellow blocks plays the role of classical Toffoli gates marked by the yellow box in figure 2 (b). Besides, those FLIP-FLOPs (e.g. **y¹** and **R¹**) marked purple act as the registers to restore the temporary data generated in the layers, and the superscripts indicate their depth positions in the circuit. Finally, the output signal of the module can be obtained from the module on the rightmost side, where Z[1:4] represents the final result $z = z_1 z_2 z_3 z_4$ of the scheme shown in figure 2(b). Obviously, this design is scalable and the structures of the corresponding circuits with larger sizes are similar.

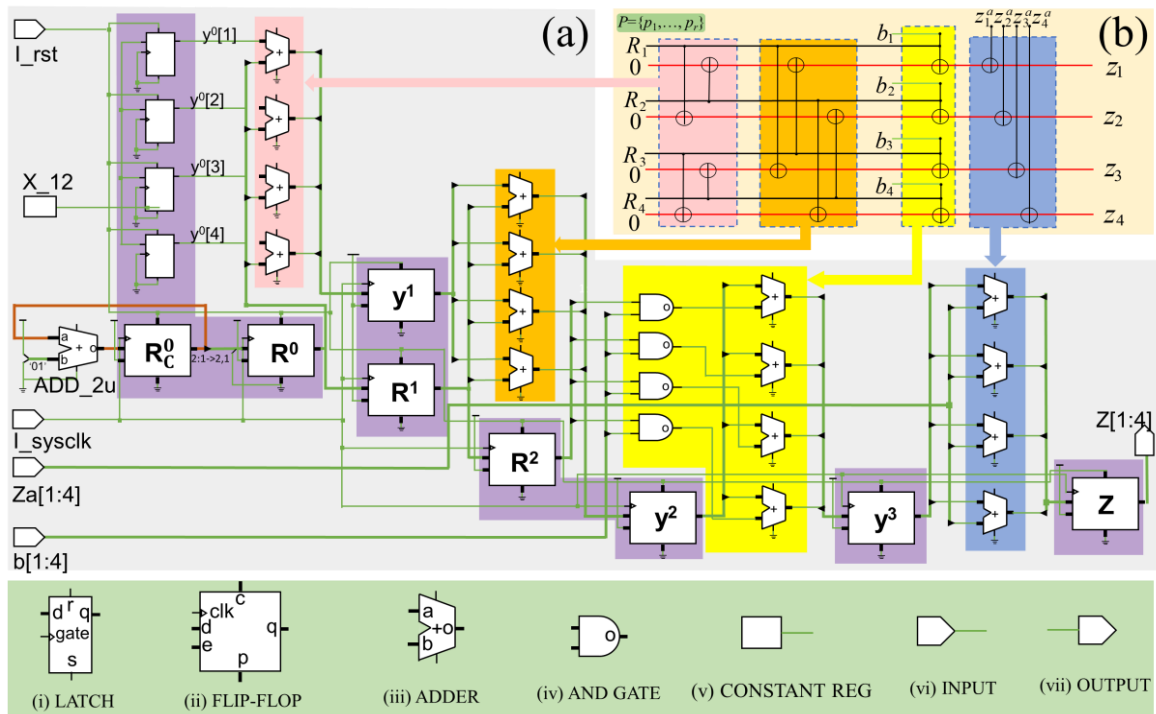

**Figure 2.** Experimental design of circuit to solve FS2D HLF problem when $n = 2 \times 2$. (a) Circuit diagram of FPGA at RTL. Four input signals of the module are marked on the leftmost side, where I_rst is the module reset signal, I_sysclk is running clock of the module, b[1:4] is a signal defining an instance of the problem, and Za[1:4] is the 4-bit output of the CLA module. The output signal Z[1:4] of the module is marked on the rightmost side. The white units represent the internal registers and logic gates of the parallel circuit, and the green solid lines represent the connections between the internal units of the module. (b) Model diagram of parallel circuit corresponding to (a), where 4 layers of parallel gate operations are marked by pink, orange, yellow and blue squares respectively. They correspond to the real circuit design of the module one by one, which are indicated by the color arrows. Insets (i)-(vii) show the names and details of digital devices used in (a), respectively.

We use Chipscope on-chip logic analyzer software to observe and record the results during the traversal of the input signals generated by ADD_2u, $\mathbf{R}_C^0$ and $\mathbf{R}^0$. For other instances, this signal generation part could be directly adapted according to their own pivot position set $P$. In figures 3(a), 3(b) and 3(c), we provide the experimental results of three instances for the case with $n = 2 \times 2$. In the experiments, to obtain a solution from the parallel circuit, only one operation clock cycle $\Delta t$ is cost. Thus, the time required to obtain all solutions of an instance is $T_F = \Delta t \cdot (\tau + 2^r)$, where $\tau$ represents the pipeline delay and $2^r$ is the number of different input strings. As a result of the fully-pipelining design, the operating clock rate is 100 MHz, i.e., $\Delta t$ is 10ns. The pipeline delay $\tau$ is related to the number of layers in parallel circuits and independent of $n$.

Figures 3(a), (b) and (c) correspond to the instances with $b = 0000$, $b = 1011$ and $b = 1111$, respectively. Each square in these graphs indicates a bit string and its color represents its statistical count (green: 0; red: 1) in the output distribution of the experiment results. The binary rank of each instance is computed as $r = 2, 3$, and 4 beforehand. Accordingly, it is seen clearly from figure 3(a) that four 4-bit strings {0000, 0110, 1001, 1111} are the solutions to the instance $b = 0000$; in figure 3(b), there are eight solutions $z = $ {0001, 0011, 0100, 0110, 1000, 1010, 1101, 1111} to $b = 1011$; in figure 3(c), there are sixteen solutions $z = $ {0000, 0001,..., 1111} to $b = 1111$. Although in principle the case $r = n$ indicates all $2^r$ basis states are solutions, we still execute the experimental process to show the feasibility of our FPGA platform.

For brevity, we would also express a $m$-bit substring being 0 or 1 as $0^{(m)}$ or $1^{(m)}$ in a $n$-bit string hereafter. For example, figures 3(d), (e) and (f) show the experimental results of three instances for $b = 0^{(16)}$, $b = 0^{(15)}1$ and $b = 0^{(14)}11$ when $n = 2 \times 2$, respectively. Note that there are too many bit strings (exactly $2^{16}$) to count, it is impossible to display all of them in the figures. Therefore, we choose four corner areas of the whole distribution to partially demonstrate the solutions and omit the rest only for the convenience of illustration. In fact, all the details of the data are obtained in our experiment rapidly. Through data processing with Matlab, it is found that there are 4096, 8192 and 16384 different solutions in figures 3(d), 3(e), and 3(f), which are consistent with their respective binary rank $r=12,13$, and 14. As for the cases with $N=3$ and 5, the experimental results are given in appendix C. These experimental statistics are verified by using open source software (e.g. ProjectQ [44]), which reflect the reliability and accuracy of our scheme. In total, the recorded runtimes of this moderate FPGA are respectively 0.04 ms, 0.08 ms and 0.16 ms for the instances with $n=4\times 4$ in figures 3(d)-(f), and 0.01 s, 0.02 s and 0.04 s for instances with $n=5\times 5$ in figures 7(a)-(c) of appendix C.

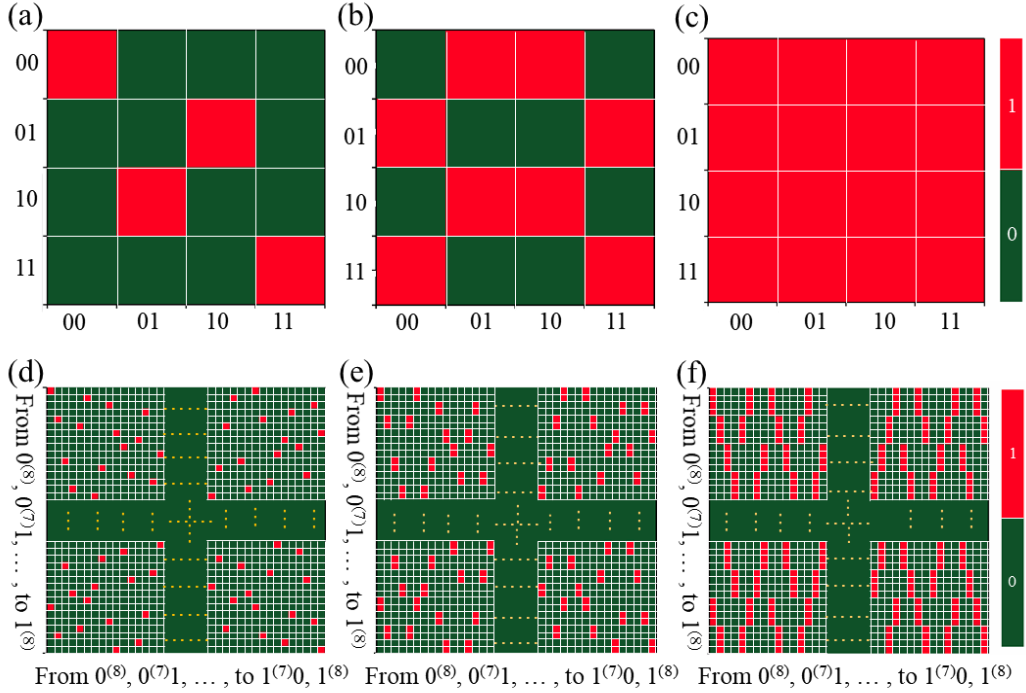

**Figure 3.** Experimental results based on FPGA to solve FS2D HLF problem, each square in the graphs indicates a bit string and its color represents the count (green: 0; red: 1). (a), (b) and (c) correspond to the cases of $n=2\times 2$ when $b=0000$, 1011 and 1111 respectively. The horizontal (vertical) coordinates of these graphs represent the first (last) two bits of each 4-bit string. (d), (e) and (f) to the cases of $n=4\times 4$ when $b=0^{(16)}$, $0^{(15)}1$ and $0^{(14)}11$, respectively. The horizontal (vertical) coordinates of these graphs respectively represent the first (last) 8 bits of each 16-bit string $\{0^{(8)}, 0^{(7)}1, \ldots, 1^{(7)}0, 1^{(8)}\}$. Only the experimental results inside areas at four corners are shown clearly for demonstration, and other results are depicted as ellipses.

## 5. Discussion and conclusion

We retrospect the performances of typical cases with all-connected 2D grids demonstrated by FPGA experiments in section 4. Such instances with $r\in[n-\sqrt{n},n]$ would own the total runtime scaling as $\Theta(2^r)$ for large $n$, and the actual time record for each run can be further optimized by increasing the operating frequency, e.g., to 233 MHz [45]. As comparison, the time of the target quantum proposal scales as $\Theta(r2^r)$, and the practical unit times for gate operation and measurement of the state-of-the-art programmable quantum experimental platform, e.g. superconducting and quantum dots systems, are tens of nanoseconds [8,41,42]. At this level, the classical-quantum time ratio defined in equation (4) can be lower than one percent for such instances with problem size $n > 100$ using a polynomial number of processors, which greatly embodies the high performance, good scalability and easy-handling of our classical scheme. As a direct application, we can utilize the ideal output distribution from our simulator together with the cross-entropy benchmarking (XEB) method [15] to verify the outcomes of an experimental implementation of specific constant-depth quantum circuits.

Besides the discussion dependent on quantum and classical hardware capacities, the design idea of the theoretical scheme itself in section 2 also has some implications for its computational (simulation) power about sampling certain quantum circuits, which can be investigated from two angles by analyzing the functions of the strings $y$ and $z^a$ in the operating process. On the one hand, we catch the specific relationship between different solutions $z$'s, i.e., the set of shift vectors $y$'s between them can be constructed by constant-depth classical circuits one by one. Therefore, the key point only lies in finding one solution $z^a$ and thus avoids the cumbersome procedures of repeatedly measuring the quantum circuit. On the other hand, the output strings $y$'s from the strictly local classical parallel circuit alone may differ from correct solutions due to the lack of quantum nonlocality. However, the quantum results can be achieved with the help of certain amount of extra information. Here for the FS2D HLF problems, the combination of $y(b,R)$ from the classical parallel circuit and $z^a(b)$ from the CLA module together gives all solutions as $y(b,R) \oplus z^a(b) = z(b,R)$. In fact, some previous studies have centered on the model of local hidden variables supplemented with additional resources (e.g. classical communication, post-selection, nonlocal boxes) to reproduce quantum outcomes [46-49], and thus our method provides a new perspective into the relationship between classical and quantum resources.

More significantly, note the derivations of solving HLF problems in section 2 are actually not restricted to the 2D grid graph. It can be verified for any given $n \times n$ symmetric binary matrix $A$ that determines a graph $G(A)$ and the quantum graph-state circuit $C_A$ with a depth $d$, the one-to-one correspondence between classical basic elements {ROU, Toffoli gate} and quantum operations {CZ gate, classically-controlled shift gate} as above introduced for 2D grid instances can be directly applied to construct a depth-$d$ classical parallel circuit plus the CLA module in figure 1(b) for simulating $C_A$. In this way the ratio in equation (4) turns into $T_C^{act} / T_Q^{act} = (c_1 \log^2 n / c_3 r 2^r d) + (c_2 / c_3 r)$, and such two-stage classical scheme for size-$n$ quantum circuits shows relatively higher performance on the instance with a longer depth $d$ and a larger binary rank $r$.

In conclusion, we have proposed a new two-stage classical computing scheme to solve the well-defined FS2D HLF problems. The key classical parallel circuit is organized in the constant depth corresponding to the quantum counterpart, supplemented with additional classical linear-algebraic algorithms. We not only analyze the time complexity of our simulator for certain instances are comparable or superior to repeating target quantum circuits, but also demonstrate its high performance by implementation on the actually designed FPGA platform. In addition to saving much running time, our scheme also possesses a good flexibility and scalability due to the maturity of the classical electronic technology at present. Furthermore, the design idea of our classical scheme as discussed in section 2 may also be generalized to broader computational problems [50,51] given the intimate correspondence between stabilizer states and graph states [32], and thus may lead to innovation in circuit designs and bring benefits for large-scale information processing.

**Acknowledgments**

This work was supported by the National key R & D Program of China under Grant No. 2017YFA0303800 and National Natural Science Foundation of China No.61421001, China Postdoctoral Science Foundation (Grant No. 2020M683049), and the Guangdong Basic and Applied Basic Research Foundation (Grant No. 2020B1515020050).

**Appendix A. Solution of the FS2D HLF problem based on quantum circuits**

We briefly review the quantum scheme that can solve the full-sampling version of the 2D HLF problem defined on a binary matrix $A$ [10]. According to the graph $G(A)$ of the square grid with $n = N \times N$ vertices as shown in figure 1(a), the corresponding quantum graph states are prepared as:

$$|\psi_{G(A)}\rangle = \prod_{(i,j)} CZ_{i,j} H^{\otimes n} |0\rangle^{\otimes n}, \tag{A1}$$

where the $CZ_{i,j}$ is the controlled-$Z$ gate acting on a pair of qubits located at two connected vertices $(i, j)$ in the graph $G(A)$ corresponding to $A_{i,j} = 1$ ($i \neq j$). Bravyi et al. have proved that by measuring each qubit $i$ of the graph state $|\psi_{G(A)}\rangle$ in the Pauli $X$ basis (for $A_{i,i} = 0$) or $Y$ basis (for $A_{i,i} = 1$), the obtained random bit string is exactly one solution denoted $z$ for the 2D HLF problem. The specific quantum scheme for performing this process is shown in figure 4(a), and we unfold its inner circuit design in figure 4(b) called $Q_N$. The input state $|0\rangle^{\otimes n}$ undergoes a layer of Hadamard gates (*H*), and then the disjointed $CZ_{i,j}$ gates appeared by $A_{i,j} = 1$ are applied for preparing the target graph state $|\psi_{G(A)}\rangle$. It can be seen from the layers 2-5 in figure 4(b) that each $CZ_{i,j}$ gate in the color of pink, green, orange, or blue is exactly in one-to-one correspondence to the connected vertices $(i, j)$ with the same color in the square graph $G(A)$ shown in figure 1(a). In the circuit design, all the *CZ* gates are arranged within at most 4 parallel layers independent of problem size *n*.

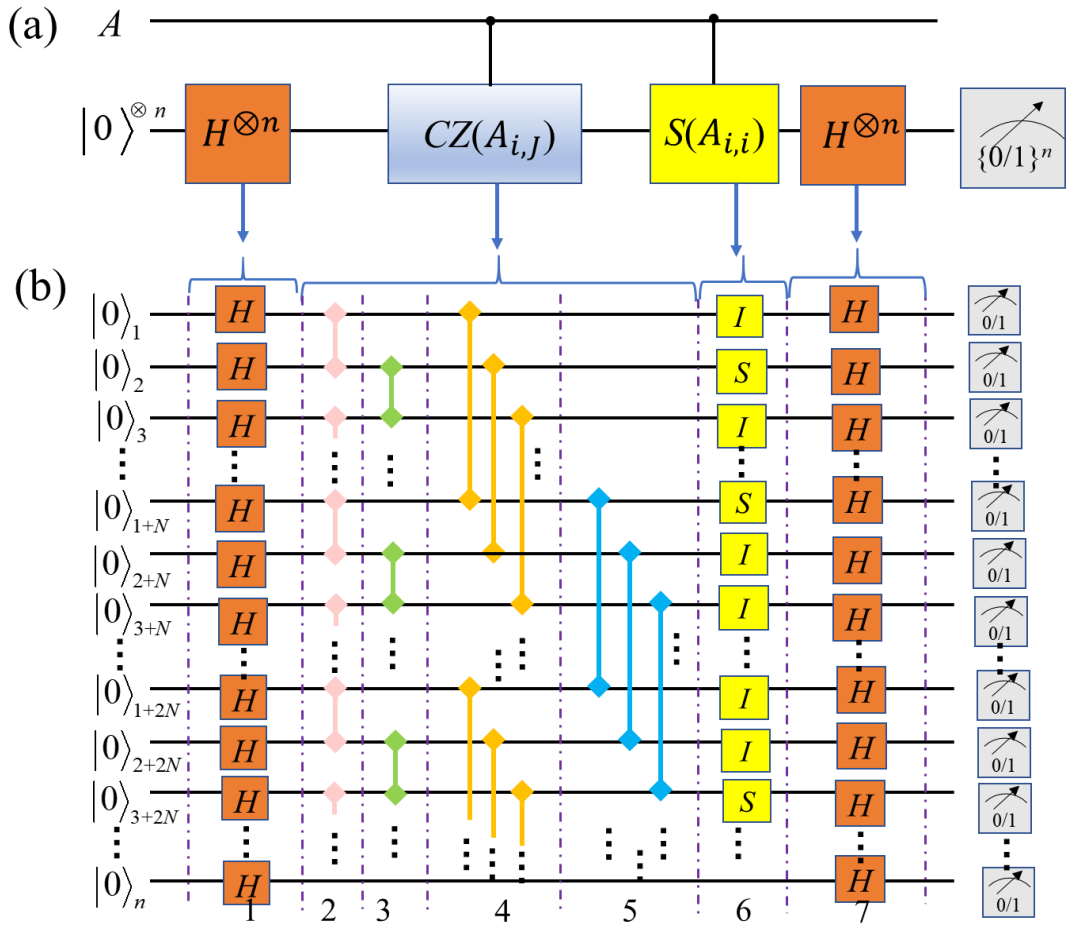

**Figure 4.** (a) The diagrammatic quantum scheme to solve FS2D HLF problem. The input includes the binary matrix *A* and a quantum basis state $|0\rangle^{\otimes n}$, which first undergoes a layer of n Hadamard gates (*H*) and becomes the state $|+\rangle^{\otimes n}$. Next, the two-qubit *CZ* gates and classically controlled phase gates (*S*) are arranged layer by layer in (b) according to graph *G(A)*. In the circuit $Q_N$, each qubit is represented by a horizontal wire, while each $CZ_{i,j}$ gate by a colored vertical segment with diamond-shaped ends connecting the qubits and corresponds to the edge $(i, j)$ with the same color in $G(A)$, and the identity gates (*I*) for $A_{i,i} = 0$ and *S* gate for $A_{i,i} = 1$ are marked by yellow squares. Finally, a layer of *H* gates are applied and the output state is measured in the computational basis. The vertical dashed lines separate the quantum circuit in (b) into different layers 1-7, indicating a circuit depth independent of input size *n*.

Subsequently, the projective measurements on this graph state in Pauli *X/Y* bases can be equivalently realized by applying identity (*I*) or phase (*S*) gate to each qubit according to its blue ($A_{i,i} = 0$) or yellow ($A_{i,i} = 1$) vertex in the square graph $G(A)$. After the final layer of Hadamard gates we get the final state

$$|\psi_q\rangle = H^{\otimes n} \prod_{1 \leq i \leq n} S_i^{A_{i,i}} |\psi_{G(A)}\rangle, \tag{A2}$$

and measure it in the standard computation basis to obtain a random result $z \in \{0,1\}^n$. Bravyi *et al.* have proved that all basis states $|z\rangle$ in the $|\psi_q\rangle$ with nonzero amplitudes exactly form the complete set of solutions to the full-sampling 2D HLF problems and are uniform distributed. In total, the depth of the quantum circuit for any large input size *n* is at most 7 as shown in figure 4(b).

## Appendix B. Proof for the binary rank of adjacency matrices of all-connected 2D grid graphs

In this part, we prove that the binary rank of the adjacency matrix of a size-*N* all-connected 2D grid graph shown in figure 1(a) must lie in the interval $[n-N, n]$ ($n = N^2$) for any diagonal string of $A$ as $b = b_1 b_2 ... b_n \in \{0,1\}^n$.

Firstly, we demonstrate the instances with all-0 diagonal elements as $b = 00...0$ exactly own the binary rank $r = n - N$, and then extend the results to general instances with $b \in \{0,1\}^n$. In the algebraic graph theory, the binary rank of an adjacency matrix is closely related to its underlying graph structure [52-56]. Therefore, here we take a simple inductive route for proof by combining both the geometric features and algebraic properties of relevant 2D grid graphs and their adjacency matrices.

We start by defining some useful concepts about the adjacency matrix $A$. $A$ can be expressed as the set of its row vectors as $A = \{r_1, r_2, ..., r_n\}$, where the $i$th row vector $r_i = [A_{i,1} \ A_{i,2} \ ... A_{i,n}]$ is determined by the connected edge of the vertex $v_i$ ($i = 1, 2, ..., n$) in the 2D grid graph $G(A)$ as shown in figure. 1(a). The aim to solve binary rank $r(A)$ of $A$ is equivalent to find the maximal linearly independent set of row vectors $L_{ind} = \{r_{p_1}, r_{p_2}, ..., r_{p_r}\}$ over the binary field GF(2). For brevity, if a given set of rows $L_r$ is linearly independent [38,39], we say the corresponding vertex set $V_r$ is independent; otherwise, if a member row vector $r_j$ equals to the linear combination of rows in $L_r$, then we call its associated vertex $v_j$ can be expressed by the vertex set $V_r$ corresponding to $L_r$ and $V_r$ is dependent. Moreover, we call a vertex set $V_r$ is "strictly dependent" if any row vector in the given $L_r$ exactly equals to the linear combination of others with all-1 coefficients. These concepts would be applied in the following, and the binary rank $r(A)$ equals to the size of maximal independent vertex subset in $G(A)$.

Obviously, a given vertex set $V_r$ must be independent if none of its subsets is strictly dependent, therefore we first consider under what conditions the target vertices can become strictly dependent. This issue can be addressed from the correspondence between the matrix elements and edges of its graph. For an adjacency matrix $A$ of a general simple undirected graph $G(A)$, we arbitrarily choose $k$ row vectors $\{r_{v_1}, r_{v_2}, ..., r_{v_k}\}$ of $A$ as target subsets with $r_{v_j} = [A_{v_j,1}, A_{v_j,2}, ..., A_{v_j,n}]$ for each $j \in [1, k]$, then the fulfillment of the strictly dependent condition

$$r_{v_1} \oplus r_{v_2} \oplus ... \oplus r_{v_k} = 0^{(n)} \tag{B1}$$

by definition implies that the row elements in the $l$th column have

$$A_{v_1,l} \oplus A_{v_2,l} \oplus ... \oplus A_{v_k,l} = 0 \ \text{for} \ l \in [1, n], \tag{B2}$$

i.e., in each column of the matrix $A$ the total number of value-1 elements with row index from $\{v_1, v_2, ..., v_k\}$ must be even. Considering the element $A_{v_j,l} = 1$ corresponds to an edge connected vertex $v_j$ and $l$, equation (B2) indicates the number of edges connected vertex $l$ and vertices numbered $\{v_1, v_2, ..., v_k\}$ must be even. In summary, the strictly dependent condition of target rows $\{r_{v_1}, r_{v_2}, \ldots, r_{v_k}\}$ can be translated into following claim of deciding the linear dependence of target vertices $\{v_1, v_2, \ldots, v_k\} \subseteq [1, n]$.

**Claim 1** *Given an $n \times n$ symmetric binary matrix A with diagonal elements $b = 0^{(n)}$ as the adjacency matrix specifying a simple undirected graph G(A), we denote a target vertex subset $V = \{v_1, v_2, \ldots, v_k\} \subseteq [1, n]$ and its connected vertex subset C in G(A), and define the degree of any vertex $l$ in C with respect to V as $\deg_V(l)$ to indicate the number of edges between $l$ and all vertices in V. Then the target vertex subset $V$ is strictly dependent if and only if $\deg_V(l)$ is even for any $l \in C$, and equivalently the row vectors $\{r_{v_1}, r_{v_2}, \ldots, r_{v_k}\}$ in A corresponding to V would add modulo 2 together to the vector $0^{(n)}$ as equation (B1) shows.*

Accordingly, figures 5(a)-(c) exhibit three typical primitive patterns of strictly dependent vertices in 2D grid graphs. For example, in figure 5(a) the target red vertex set $V = \{v_1, v_2, \ldots, v_k\}$ are arranged in a diagonal skew line and their connected vertices are marked green. The degree of each green vertex with respect to target red vertex set $V$ is 2 (the circled number) and thus the red vertex set $V = \{v_1, v_2, \ldots, v_k\}$ are strictly dependent according to **Claim 1**. This result also holds for the red vertices located in the horizontal and vertical lines of figure 5(b), and the strictly dependent red vertices shown in figure 5(c) cover the patterns in figures 5(a) and 5(b).

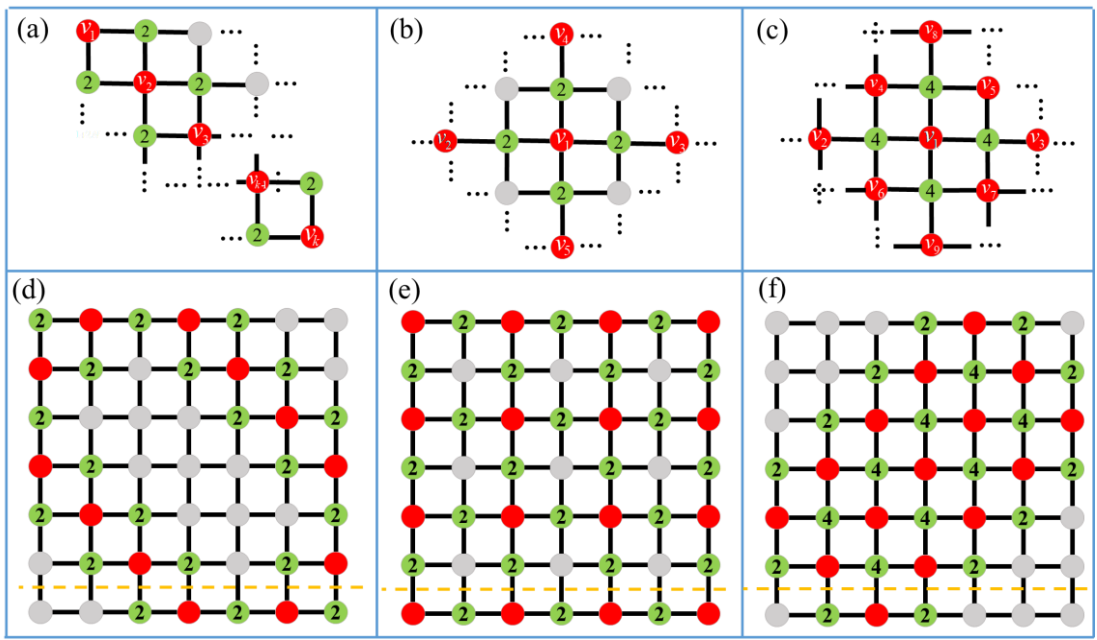

**Figure 5.** For a symmetric binary matrix with all-0 diagonal elements $A$, (a)-(c) show three typical primitive patterns for strictly dependent vertex subsets marked red in the $n=N \times N$ 2D grid graph $G(A)$, the vertices connected to red ones are marked green and the rest marked gray. The number 2 or 4 in a green vertex indicates its degree with respect to the red vertices. (d), (e) and (f) are complete $7 \times 7$ examples of (a), (b) and (c), respectively, and the orange dashed line separates all $n$ vertices into two subsets $\{v_1, v_2, \ldots, v_{n-N}\}$ and $\{v_{n-N+1}, \ldots, v_n\}$. The row vectors in $A$ corresponding to $\{v_1, v_2, \ldots, v_{n-N}\}$ can be proved to be the maximal linearly independent subset.

Figures 5(d), (e) and (f) are respectively $7 \times 7$ concrete examples of figures 5(a), (b) and (c), where the target subset $V$ marked red are strictly dependent, its connected subset $C$ marked green has even $\deg_V(l \in C) = 2$ or 4 indicated by the circled number, and other vertices marked gray. The strictly dependent red vertices in these examples involve those located at margins or corners of the 2D grid graph. The orange dashed lines in figures 5(d) (e) and (f) are marked to separate all $n = N \times N$ vertices into two subsets denoted $V_{\text{row}(1 \to N-1)} = \{v_1, v_2, \ldots, v_{n-N}\}$ and $V_{\text{bottom}} = \{v_{n-N+1}, \ldots, v_n\}$. At this point, we can prove $V_{\text{row}(1 \to N-1)}$ is a maximal independent vertex subset by using **Claim 1** in an iterative way. Firstly, we demonstrate that no strictly dependent subsets of $V_{\text{row}(1 \to N-1)}$ exist by checking its vertices row by row from bottom to top. For the target vertices $V_{(N-1)} = \{v_{n-2N+1}, \ldots, v_{n-N}\} \subseteq V_{\text{row}(1 \to N-1)}$ located at the $(N-1)$th row in the 2D grid graph, we notice that for each vertex $v_t \in V_{(N-1)}$ and any vertex subset $V_t$ satisfying $v_t \in V_t \subseteq V_{\text{row}(1 \to N-1)}$, the vertex denoted $v_{t+N}$ below $v_t$ belongs to $V_{\text{bottom}}$ and thus $\deg_{V_t}(v_{t+N}) = 1$. This violation of **Claim 1** indicates that in $V_{\text{row}(1 \to N-1)}$ any subset including vertices in $V_{(N-1)}$ cannot be strictly dependent. That is, vertices in $V_{(N-1)}$ are excluded from any possible strictly dependent subsets of $V_{\text{row}(1 \to N-1)}$. In a similar way, it can be derived that the vertices $V_{(N-2)} = \{v_{n-3N+1}, \ldots, v_{n-2N}\}$ located at the $(N-2)$th row in the 2D grids are excluded from any possible strictly dependent subsets of $V_{\text{row}(1 \to N-2)} = \{v_1, v_2, \ldots, v_{n-2N}\}$. Repeating this derivation $N-1$ times from vertices located at the $(N-1)$th row to the first (topmost) row, it can be concluded that none of these vertices can form a strictly dependent subset. Furthermore, $V_{\text{row}(1 \to N-1)} \cup \{v_k\}$ for any $v_k \in V_{\text{bottom}}$ is dependent as the strictly dependent pattern (i.e. red circles) in figure 5(c) tells, with figure 5(f) being an explicit

example. Therefore, we demonstrate that for the instances with $b = 0^{(n)}$, the row vectors numbered $\{1, 2, ..., n-N\}$ form one maximal linearly dependent set of $A$ and its binary rank is $r(A) = n - N$.

Now we proceed to deal with the binary rank of general instances with any diagonal elements $b \in \{0,1\}^n$ ($b_i = A_{i,i}$). In the 2D grid graph, we mark the $i$th vertex corresponding to diagonal element $b_i = A_{i,i} = 1$ with a self-loop to connect to itself. Thus for a given target vertex subset $V$, its connected vertex subset $C$ may be expanded compared to $b = 0^{(n)}$ due to self-loops, and the **Claim 1** about strictly dependent subsets can be adapted as the following generalized claim.

**Claim 2** *A given $n \times n$ symmetric binary matrix A with diagonal elements $b \in \{0,1\}^n$ can uniquely specify an undirected graph G(A) such that the off-diagonal part of A determines the edges between different vertices and any bit $b_j = 1$ in $b = b_1 b_2 ... b_n$ would add a self-loop to the vertex $j \in [1, n]$, i.e., an edge that connects this vertex to itself. We denote a target vertex subset $V = \{v_1, v_2, ..., v_k\} \subseteq [1, n]$ and its connected vertex subset C in G(A), and define the degree of any vertex l in C with respect to V as $\deg_V(l)$ to indicate the number of edges between l and all vertices in V. Note the vertices that belong to V and have self-loops must belong to C. Then the target vertex subset V is strictly dependent if and only if $\deg_V(l)$ is even for any $l \in C$, and equivalently the row vectors $\{r_{v_1}, r_{v_2}, ..., r_{v_k}\}$ in A corresponding to V would add modulo 2 together to the vector $0^{(n)}$ as equation. (B1).*

By **Claim 2** we can demonstrate the target vertex subset $V_{\text{row}(1 \to N-1)} = \{v_1, v_2, ..., v_{n-N}\}$ is still independent regardless of any possible self-loops of certain vertices in the all-connected 2D graph. The method of iteratively checking the independence of vertices located at rows indexed $N-1$, $N-2$, ..., 1 for proof is exactly the same as that for illustrating the instance with $b = 0^{(n)}$. Note here the vertex subset $V_{\text{row}(1 \to N-1)}$ may not be a maximal independent one for instances other than $b = 0^{(n)}$, we therefore conclude that the binary rank of any instance with all-connected 2D grids and $b \in \{0,1\}^n$ is $n - N \leq r(A) \leq n$.

In summary, we prove the binary rank of a given $n \times n$ symmetric binary matrix $A$ restricted to an $N \times N$ all-connected 2D grid graph of size $n$ ($n = N^2$) must be $r \in [n - N, n]$. The aim to find the maximal linearly independent subset of row vectors in $A$ is neatly addressed from the observation on the geometric structure (e.g. the connected edges and self-loops) of the 2D graph, which is an intuitive way to understand. Besides these instances, we declare the **Claim 2** about deciding strictly dependent row vectors actually holds for any binary symmetric matrix.

**Appendix C. More experimental results for FS2D HLF problem**

As introduced in section 4, we employ customized FPGA platform to implement the classical parallel scheme for solving FS2D HLF problem, and present the statistical distribution of experimental results for several instances with size $n = 2 \times 2$ and $n = 4 \times 4$ in figure. 3. In a general way, problem instances with other sizes can also be solved with similarly designed FPGA architecture. For simplicity, here we record the $n$-bit string in the form $b = 00\ldots0$ as $b = 0^{(n)}$.

Figures 6(a), (b) and (c) show the experimental results of three instances for $b = 0^{(9)}$, $b = 0^{(8)}1$ and $b = 0^{(7)}11$ at $n = 3 \times 3$, respectively. It can be counted from figures 6(a), 6(b) and 6(c) that there are totally 64, 128, and 256 solutions for these three instance, which coincide with their respective binary rank $r = 6, 7,$ and 8.

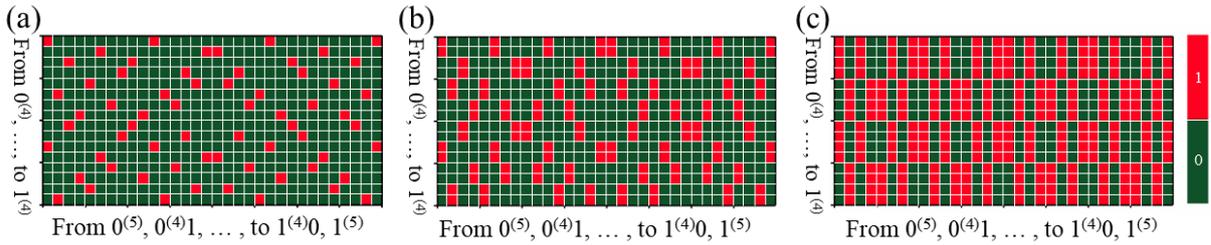

**Figure 6.** Experimental results based on FPGA to solve FS2D HLF problem with size $n = 3 \times 3$, each square in the graphs indicates a bit string and its color represents the statistical count (green: 0; red: 1). (a), (b) and (c) correspond to the cases with $b = 0^{(9)}$, $b = 0^{(8)}1$ and $b = 0^{(7)}11$, respectively. The horizontal and vertical coordinates of these graphs respectively represent the first 5 bits $\{00000, 00001, \ldots, 11111\}$ and last 4 bits $\{0000, 0001, \ldots, 1111\}$ of each 9-bit string.

Figures 7(a), (b) and (c) show the experimental results of three instances for $b = 0^{(25)}$, $b = 0^{(24)}1$ and $b = 0^{(23)}11$ at $n = 5 \times 5$, respectively. Note there are too many bit strings ($2^{25}$) to count, it is impossible to display all of them in the figures. The experimental results are only shown clearly inside sampled areas at four corners, each of which contains the counts of $7 \times 14$ bit strings as representatives. In other areas the counts of bit strings are just shown as ellipses. Through data

processing by Matlab, it is found that that there are totally $2^{20}$, $2^{21}$, and $2^{22}$ different solutions for these three instances as shown in figures 7(a), 7(b) and 7(c), respectively, which coincide with their respective binary rank $r = 20, 21,$ and $22$.

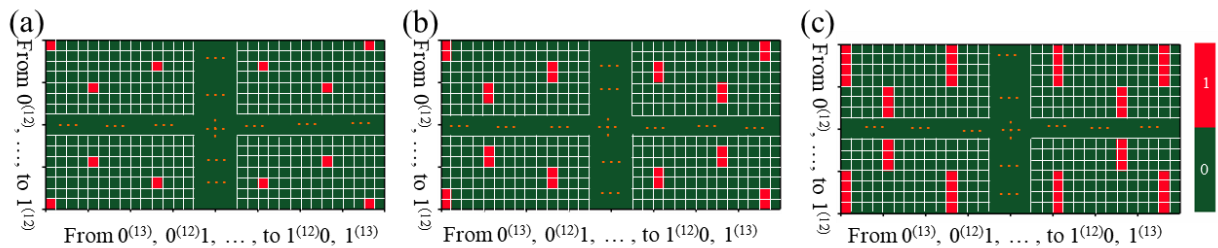

**Figure 7.** Experimental results based on FPGA to solve FS2D HLF problem with size $n = 5 \times 5$, each square in the graphs indicates a bit string and its color represents the statistical count (green: 0; red: 1). (a), (b) and (c) correspond to the cases with $b = 0^{(25)}$, $b = 0^{(24)}1$ and $b = 0^{(23)}11$, respectively. The horizontal and vertical coordinates of these graphs respectively represent the first 13 bits { $0^{(13)}$, $0^{(12)}1$ ,…, $1^{(13)}$ } and last 12 bits { $0^{(12)}$, $0^{(11)}1$ ,…, $1^{(12)}$ } of each 25-bit string. Only the experimental results inside areas at four corners are shown clearly for demonstration, and other results are depicted as ellipses.

# References

[1] Deutsch D 1985 Quantum theory, the Church-Turing principle and the universal quantum computer *Proc. R. Soc. A* **400** 97–117
[2] Shor P W 1997 Polynomial-time algorithms for prime factorization and discrete logarithms on a quantum computer *SIAM J. Comput.* **26** 1484
[3] Grover L K 1997 Quantum mechanics helps in searching for a needle in haystack *Phys. Rev. Lett.* **79** 325
[4] Nielsen M A and Chuang I L 2011 *Quantum Computation and Quantum Information: 10th Anniversary Edition* 10th edn (New York: Cambridge University Press)
[5] Georgescu I M, Ashhab S and Nori F 2014 Quantum simulation *Rev. Mod. Phys.* **86**, 153
[6] Harrow A W and Montanaro A 2017 Quantum computational supremacy *Nature* **549**, 203
[7] Preskill J 2012 Quantum computing and the entanglement frontier arXiv:1203.5813 [quant-ph]
[8] Arute F *et al* 2019 Quantum supremacy using a programmable superconducting processor *Nature* **574** 505–510
[9] Zhong H.-S *et al* 2020 Quantum computational advantage using photons *Science* **370**, 1460-1463
[10] Bravyi S, Gosset D, König R, 2018 Quantum advantage with shallow circuits *Science* **362**, 308-311
[11] Aaronson S and Arkhipov A 2010 The computational complexity of linear optics *Proc. of the 43rd Annual ACM Symp. on Theory of Computing* (New York: ACM) pp 333–342
[12] Lund A P, Bremner M J and Ralph T C 2017 Quantum sampling problems, BosonSampling and quantum supremacy *NPJ Quantum Inf.* **3** 15
[13] Bremner M J, Montanaro A and Shepherd D J 2016 Average-case complexity versus approximate simulation of commuting quantum computations *Phys. Rev. Lett.* **117** 080501
[14] Fujii K and Morimae T 2017 Commuting quantum circuits and complexity of Ising partition functions *New J. Phys.* **19** 033003

[15] Boixo S, Isakov S V, Smelyanskiy V N, Babbush R, Ding N, Jiang Z, Bremner M J, Martinis J M and Neven H 2018 Characterizing quantum supremacy in near-term devices *Nat. Phys.* **14** 1–6

[16] Chen M-C, Li R, Gan L, Zhu X, Yang G, Lu C-Y and Pan J-W 2020 Quantum teleportation-inspired algorithm for sampling large random quantum circuits *Phys. Rev. Lett.* **124** 080502

[17] Gao X, Wang S-T and Duan L M 2017 Quantum Supremacy for Simulating a Translation-Invariant Ising Spin Model *Phys. Rev. Lett.* **118** 040502

[18] M.-H. Yung 2019 Quantum supremacy: some fundamental concepts *Natl. Sci. Rev.* **6**, 22

[19] Chen Z Y, Zhou Q, Xue C, Yang X, Guo G C and Guo G P 2018 64-qubit quantum circuit simulation *Sci. Bull.* **63** 964-71

[20] Villalonga B *et al* 2020 Establishing the quantum supremacy frontier with a 281 Pflop/s simulation *Quantum Sci. Technol.* **5** 034003

[21] Villalonga B, Boixo S, Nelson B, Henze C, Rieffel E, Biswas R and Mandrà S 2019 A flexible high-performance simulator for verifying and benchmarking quantum circuits implemented on real hardware *npj Quantum Inf.* **5** 1–16

[22] Neville A, Sparrow C, Zlifford R, Johnston E, Birchall P M, Montanaro A and Laing A 2017 Classical boson sampling algorithms with superior performance to near-term experiments *Nat. Phys.* **13** 1153–7

[23] Wu J, Liu Y, Zhang B, Jin X, Wang Y, Wang H and Yang X 2018 A benchmark test of boson sampling on Tianhe-2 supercomputer *Natl Sci. Rev.* **5** 715

[24] Alexandra E Moylett *et al* 2020 Classically simulating near-term partially-distinguishable and lossy boson sampling *Quantum Sci. Technol.* **5** 015001

[25] Linke N M *et al* 2017 Experimental comparison of two quantum computing architectures *Proc. Natl. Acad. Sci.* **114**, 3305-3310

[26] Choo K, Von Keyserlingk C W, Regnault N, Neupert T, 2018 Measurement of the entanglement spectrum of a symmetry-protected topological state using the IBM quantum computer *Phys. Rev. Lett.* **121** 086808

[27] Gong M *et al* 2019 Genuine 12-qubit entanglement on a superconducting quantum processor *Phys. Rev. Lett.* **122**, 110501

[28] Omran A *et al* 2019 Generation and manipulation of Schrödinger cat states in Rydberg atom arrays *Science* **365**, 570-574

[29] C. Song *et al* 2019 Generation of multicomponent atomic Schrödinger cat states of up to 20 qubits *Science* **365** 574-577

[30] Gottesman D 1998 The Heisenberg representation of quantum computers arXiv 9807006 [quant-ph]

[31] Aaronson S, Gottesman D 2004 Improved simulation of stabilizer circuits *Phys. Rev. A* **70** 052328

[32] Anders S, Briegel H J 2006 Fast simulation of stabilizer circuits using a graph-state representation *Phys. Rev. A* **73** 022334.

[33] Nest M 2008 Classical simulation of quantum computation, the Gottesman-Knill theorem, and slightly beyond arXiv:0811.0898 [quant-ph]

[34] Gosset D, Grier D, Kerzner A, Schaeffer L 2020 Fast simulation of planar Clifford circuits arXiv:2009.03218 [quant-ph]

[35] Arunachalam S, Belovs A, Childs A M, *et al* 2020 Quantum Coupon Collector *15th Conference on the Theory of Quantum Computation, Communication and Cryptography* (TQC 2020, arXiv: 2002.07688)

[36] Borodin A, von zur Gathem J, Hopcroft J 1982 Fast parallel matrix and GCD computations *23rd Annual Symposium on Foundations of Computer Science* IEEE, pp 65-71.

[37] Mulmuley K 1986 A fast parallel algorithm to compute the rank of a matrix over an arbitrary field *Proc. of the 18th Annual ACM Symp. on Theory of Computing* pp. 338-339

[38] Roman S, Axler S, Gehring F W, 2005 *Advanced Linear Algebra* (New York: Springer)

[39] Hogben L, 2013 *Handbook of linear algebra* 2nd edn (CRC press)

[40] Johansson N, Larsson J A 2019 Quantum simulation logic, oracles, and the quantum advantage *Entropy* **21**, 800

[41] Krantz P *et al* 2019 A quantum engineer's guide to superconducting qubits, *Appl. Phys. Rev.* **6**, 021318

[42] Hendrickx N W, Franke D P, Sammak A, Scappucci G, and Veldhorst M 2020 Fast two-qubit logic with holes in germanium, *Nature* **577**, 487

[43] Wei K X *et al* 2020 Verifying multipartite entangled Greenberger-Horne-Zeilinger states via multiple quantum coherences *Phys. Rev. A* **101** 032343

[44] Steiger D S, Häner T, Troyer M 2018 ProjectQ: an open source software framework for quantum computing *Quantum* **2**, 49

[45] Mahmud N, El-Araby E, Caliga D 2019 Scaling reconfigurable emulation of quantum algorithms at high precision and high throughput *Quantum Engineering* **1**, e19

[46] Toner B F, Bacon D 2003 Communication cost of simulating Bell correlations *Phys. Rev. Lett.* **91** 187904

[47] Degorre J, Laplante S, Roland J 2005 Simulating quantum correlations as a distributed sampling problem *Phys. Rev. A* **72** 062314

[48] Barrett J, Pironio S 2005 Popescu-Rohrlich correlations as a unit of nonlocality. *Phys. Rev. Lett.* **95**, 140401

[49] Buhrman H, Cleve R, Massar S and de Wolf R 2010 Nonlocality and communication complexity *Rev. Mod. Phys.* **82** 665

[50] Watts A B, Kothari R, Schaeffer L, and Tal A 2019 Exponential separation between shallow quantum circuits and unbounded fan-in shallow classical circuits *Proc. of the 51st Annual ACM SIGACT Symp. on Theory of Computing* (New York: ACM) pp. 515-526

[51] Bravyi S, Gosset D, Kœnig R, and Tomamichel M 2020 *Nat. Phys.* **16**,1040

[52] Godsil C and Royle G F, 2001 *Algebraic Graph Theory* Springer-Verlag, New York

[53] Brouwer A E, Van Eijl C A 1992 On the p-rank of the adjacency matrices of strongly regular graphs *J. Algebr. Comb.* **1** 329–346

[54] Peeters R 2002 On the p-ranks of the adjacency matrices of distance-regular graphs *J. Algebr. Comb.* **15** 127-149

[55] Pflueger N 2011 Graph reductions, binary rank, and pivots in gene assembly *Discrete Appl. Math.* **159** 2117-2134

[56] Abiad A, Haemers W H 2016 Switched symplectic graphs and their 2-ranks, *Des. Codes Cryptogr.* **81** 35-41